\documentclass[aps,pre,twocolumn,groupedaddress,showpacs,preprintnumbers,amsmath,amssymb,floatfix]{revtex4}
\usepackage{graphicx}
\usepackage{bm}

\begin{document}

\title{Interplay of chemotaxis and chemokinesis 
mechanisms in bacterial dynamics}

\author{Maria R. D'Orsogna$^{1}$, Marc Suchard$^{2}$, and Tom Chou$^{2}$}
\affiliation{$^{1}$Department of Physics \& Astronomy, 
University of California, Los Angeles, CA
90095-1547}
\affiliation{$^2$Department of Biomathematics, 
University of California, Los Angeles, CA
90095-1766} 

\date{\today}

\begin{abstract}
Motivated by observations of the dynamics of {\it Myxococcus xanthus}, 
we present a self-interacting random walk
model that describes the competition between chemokinesis and chemotaxis.
Cells are constrained to move in one dimension, but release a chemical 
chemoattractant at a steady state.  
The bacteria senses the chemical that it produces.  The probability of
direction reversals is modeled as a function of both the absolute level 
of chemoattractant sensed directly under each cell as well as the 
gradient sensed across the length of the cell. 
If the chemical does not degrade or diffuse rapidly, 
the one dimensional trajectory depends on
the entire past history of the trajectory. 
We derive the corresponding Fokker-Planck
equations, use an iterative mean field approach that we solve 
numerically for short times, and
perform extensive Monte-Carlo simulations of the model.  
Cell positional distributions and the
associated moments are computed in this feedback system. 
Average drift and mean squared
displacements are found.
Crossover behavior among different diffusion regimes are found.

\end{abstract}


\maketitle

\makeatletter
\global\@specialpagefalse 
\def\@oddhead{{\it Physical Review E}, {\bf 68}, 021925, 2003\hfill}
\let\@evenhead\@oddhead
\makeatother      

\par

\section{Introduction}
\label{sec:introduction}

The dynamics and pattern formation of bacteria serve as paradigms
in understanding many properties of multicellular interacting systems, 
such as collective behavior, self organization, evolution, and development 
\cite{preston, levine}.
The characteristics of bacterial motility and aggregation 
depend on numerous biological and chemical parameters,
which include light exposure, temperature, concentration of food or of
other substances, 
and lead to the existence of many classes of cell motion.
Since a bacterium cannot utilize electromagnetic or acoustic
radiation to sense its environment, it must rely on physical
contact and its motility may depend on the production,
diffusion and degradation of a relevant set of sensed chemicals.
These chemicals are called chemoattractants or chemorepellants if they 
attract or repel bacteria, respectively.
Examples of chemoattractants include food - sugars and amino acids -
whereas antiobiotics, fatty acids or
other noxious substances are chemorepellant.

A prototype and well studied example of bacterial motility
is the run and tumble mechanism used by {\it Escherichia coli}
in liquid environments.
The specialized structures allowing for cellular motion are known as 
flagellae, spinning helical tails which 
extend from the cellular membrane
into the surrounding environment:
counter-clockwise rotation of the flagellar motor leads to 
a one-directional run whereas clockwise rotation
causes tumbling along a random direction.
During its run,
an {\it E.coli} cell periodically senses
for chemosensitive substances, and,
by comparing concentration levels
in new and old environments, adjusts its motion 
and its likelihood to tumble in a new, randomly chosen, direction
\cite{berg,woodward}.
If the cell is moving in the direction of increasing
attractant, for instance, the probability of tumbling is lowered,
so that the run time and `mean free path' length in this 
direction are increased. 
An {\it E.coli} cell, thus,
compares chemical concentrations at nearby points, so that
its future motion is determined by a chemical gradient \cite{bergold,brenner}
via the mechanism of {\it chemotaxis}.

Chemotaxis may occur in other non-acqueous systems, 
such as in colonies of the bacterium {\it Myxococcus xanthus}
or of the eukaryote {\it Dictyostelium discoideum}
(slime-mould)
which crawl on agar plates \cite{kearns}.
The exact regulatory mechanism leading to the surface
gliding of these 
rod-like and slow moving organisms is yet unknown,
and responses to external stimuli appear to be more complex than in 
enteric bacteria such as ${\it E. coli}$ \cite{rappel}.
The cells undergo chemotaxis, but lack, or show very modest,
responses to specific nutrient stimuli
\cite{shi, tieman}, suggesting that the 
chemotactic behavior is due mainly to self-generated signaling chemicals
\cite{kearns2}.
The fact that cell density regulates typical reversal frequencies 
may be a consequence of one bacteria sensing the higher levels of 
signaling chemical 
in the presence of others \cite{Shi96}.
In particular, under starving conditions, the cells are
driven by auto-released fibril trails to aggregate 
and form multicellular compounds called fruiting bodies.
These compounds raise above the agar plate and eventually sporulate into  
new and more resistant organisms which are then 
released to the environment in search of new sources of nutrients.
 
One major difference between these systems and enteric bacteria
is that the chemoattractants embedded in or self-deposited 
on the agar plates are 
relatively immobile on the time scale of bacterial motility,
whereas chemoattractant
substances in acqueous environments
have a finite diffusion constant.
Like {\it E.coli}, the gliding cells can only sense what they
come into immediate contact with, and since
the chemosensitive substance diffuses slowly, large fibril
concentration differences exist over the length of the cell.
The gradient driving the chemotaxis mechanism is 
then determined between positions within a cell body length,
as opposed to the large distances involved in 
enteric bacteria chemotaxis.

Another possible mechanism driving cellular motility is
{\it chemokinesis}. 
Here, changes in the direction of motion or in the cell velocity
are determined locally, without recourse to the determination
of concentration gradients.
Chemokinesis may be a function of temperature, substrate adhesion,
salt concentration or substances
affecting the internal metabolism of the cell,
slowing or enhancing its motion or turning frequency \cite{soll}.

The observed patterns of movement 
of {\it M. xanthus} and of slime-mould
suggest that the motility of these systems
may be affected by self-induced chemokinesis as well as self-induced
chemotaxis. {\it M. xanthus} move by shooting out and retracting pili from the
ends of their prolate bodies \cite{spormann, Mer00}.
In nutritionally abundant environments or under conditions
of high population density, both cellular systems 
generate repulsive fibril trails,
leading to cell dispersion and the colonization of
new regions of the agar plate. The mechanism is hypothesized
to be dictated by
chemokinesis \cite{zusman,zusman2}.

Previous theoretical studies into the collective dynamics of
interacting motile bacteria, both of {\it E. coli} and {\it M. xanthus}
type,  
include coarse-grained
convective-diffusive transport type models \cite{Alt,Kel74,Dal98,Pal00,Nos76}
and lattice-type simulations \cite{Oth97,hillen,Ste95}. 
These studies have 
considered the effects of interactions with concentration fields,
modeling chemotaxis or chemokinesis or, as in Ref.\,\cite{rohde}, both
chemosensitive mechanisms. 
The continuum theories take into account turning rates, and changes
in speed or direction of motion,
however they
generally ignore long time path history effects induced by the 
sensed chemicals. In these systems, the bias to the motion arises from
an external, fixed stimulus and not a dynamical one.
Similar studies - with a fixed external trail -  have been presented in 
different contexts, such as the aggregation of
ants due to chemotaxis \cite{ants}. 
Other authors \cite{maritan} have presented systems in which the fibril
trail that dictates the motion of the cells is
determined over an extended area and is not sensed locally.

The aim of this paper is to model and disentangle
the effects of {\it self-generated} chemotaxis and 
chemokinesis, acting individually or concurrently
and under diverse attractrant and repulsive conditions. 
In contrast to many of the previous studies,
the chemical trail
of our systems will emerge from the motion of the bacterial cells and
evolve with the cellular motion itself.
Observations of individual organism paths may be 
more revealing of cellular dynamics
than multicellular ensembles, and we restrict our attention
to the behavior of a single bacterium.
Even if isolated, the dynamics of this single cell
will be affected by its self-secreted
fibril trail, and the resulting pattern will be that of a 
self-interacting random walker. 
We shall try to understand cellular motion
by developing, and analytically solving, the equations of motion for 
an isolated {\it M. xanthus} cell under a mean-field approximation, and
by means of simple Monte-Carlo simulations, 
incorporating chemotaxis and chemokinesis to various degrees.
The underlying assumption of this paper is that the
chemosensitive material acts as a
regulator of cell motion affecting only the frequency of 
direction reversal, via chemokinesis or chemotaxis.
The speed of the bacterium is assumed constant 
in either direction and is controlled by inherent cellular
bio-energetics.

Experimental single cell trajectories show that an isolated cell
travels along its axis, occasionally veering off,
and for the sake of simplicity, we will only consider one-dimensional 
dynamics. In order to validate our models, 
cell particle tracking can be eventually performed 
in one-dimensional etchings or imprints into the agar
plates.

\section{Constant reversal rates}

In this Section we derive a general Fokker-Planck
equation for the single one-dimensional random walker. 
As discussed in the Introduction, we assume that 
the `bacterium', 
or particle, travels with constant 
speed $v_0$, and is subject to an {\it ad hoc}
directional reverse mechanism that incorporates both 
chemotaxis and chemokinesis. To start,
we ignore the effect of both phenomena, 
and we simply investigate the role of a constant reversal rate.

\begin{figure}
\includegraphics[height=1in]{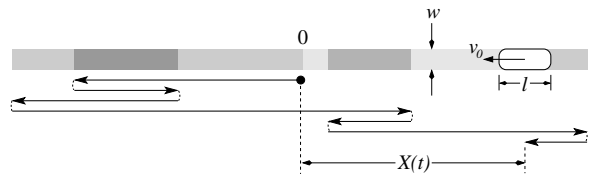}
\caption{The trajectory of a single bacterium of width $w$ and length 
$\ell$ crawling in one
dimension and depositing chemoattractant (or repellent) on the surface. 
The density of shading of the footprint is proportional to the 
total attractant deposited and arises from the
hypothetical trajectory depicted by the arrows beneath. 
For the sake of simplicity, we have
neglected the cell length $\ell$ in the depiction of the trajectory.}
\label{1D}
\end{figure}

The probability that a bacterium is centered
at position $x$ at time $t$ with velocity $\pm v_0$
will be denoted by $P_{\pm}(x,t)$ and
the probability of a directional switch in time $dt$ is 
$dt/ \tau = \gamma ~ dt$. 
If $\gamma$ is independent of $(x,t)$ we can write:

\begin{eqnarray}
\label{proa}
P_+(x,t) &=& P_+(x-v_0 dt, t - dt) (1-\gamma_{+} dt) + \\ 
\nonumber
&& P_-(x,t) \gamma_{-} dt,
\\
\label{prob}
P_-(x,t) &=& P_-(x+v_0 dt, t - dt) (1-\gamma_{-} dt) + \\
\nonumber
&& P_+(x,t) \gamma_{+} dt.
\end{eqnarray}

\noindent
In the above equations, we have included the possibility that
$\gamma$ depends on the type of reversal, since in general the
reversal rates from one direction to the other are not
the same. Specifically:
\begin{eqnarray}
\gamma_{+}: +v_0 \rightarrow -v_0, \qquad
\gamma_{-}: -v_0 \rightarrow +v_0.
\end{eqnarray}

\noindent
Expanding Eqns.\,(\ref{proa}) and (\ref{prob}) and keeping only the
$O (dt)$ terms, we find:
\begin{eqnarray}
\label{pro1}
\dot P_+(x,t)+v_0 \partial_x P_+(x,t) = -\gamma_+ P_+(x,t)
+\gamma_- P_-(x,t), \\
\label{pro2}
\dot P_-(x,t)-v_0 \partial_x P_-(x,t) = -\gamma_- P_-(x,t)
+\gamma_+ P_+(x,t). 
\end{eqnarray}

\noindent
The initial conditions are chosen so that at $t=0$ the particle
is localized at $x=0$ with amplitudes $a_\pm$
satisfying the condition that $a_++a_-=1$:
\begin{equation}
\label{initial}
P_+(x,0)= a_+ \delta(x), \qquad P_-(x,0)= a_- \delta(x),
\end{equation}

\noindent
We obtain:
\begin{eqnarray}
\label{pplus}
P_+(x,t)&=& 
\exp \left[{\frac{(q_-x-q_+v_0t)}{2 v_0}}\right] \\
\nonumber
&&
\hspace{-1cm}
\left\{
\frac{a_+}{2}
[\delta(x-v_0t)+\delta(x+v_0t)]
+
\Theta[v_0t - |x|] \right. \\
&&
\nonumber
\hspace{-1cm}
\left[\frac{a_+ Q q_+}{4 v_0}
\sqrt{\frac{v_0 t+x}{v_0 t -x}}
~ I_1 \left(\frac{q_+ Q}{2v_0} 
\sqrt{(v_0t)^2-x^2}\right)  \right. \\
&&
\hspace{-1cm}
\nonumber
+\left. \left. \frac{a_- \gamma_-}{2 v_0}
I_0 \left( \frac{q_+ Q}{2 v_0} \sqrt{(v_0t)^2 -x^2}\right)
\right] 
\right\};
\end{eqnarray}

\begin{eqnarray}
\label{pminus}
P_-(x,t) &=&  
\exp \left[{\frac{(q_-x-q_+v_0t)}{2 v_0}}\right] \\
\nonumber
&& \hspace{-1cm} \left\{
\frac{a_-}{2}
[\delta(x-v_0t)+\delta(x+v_0t)] +
\Theta[v_0t - |x|] \right. \\
&&
\nonumber
\hspace{-1cm}
\left[\frac{a_- Q q_+}{4 v_0}
\sqrt{\frac{v_0 t-x}{v_0 t +x}}
~ I_1 \left(\frac{q_+ Q}{2v_0} 
\sqrt{(v_0t)^2-x^2}\right) \right. \\
\nonumber
&& \left. \left. 
\hspace{-1cm} + \frac{a_+ \gamma_+}{2 v_0}
I_0 \left( \frac{q_+ Q}{2 v_0} \sqrt{(v_0t)^2 -x^2}\right)
\right] 
\right\}.
\\
\nonumber
\end{eqnarray}

\noindent
Here,
the parameters $q_+$, $q_-$ and $Q$ are defined so 
that $q_+=(\gamma_+ +\gamma_-)$, $q_-=(\gamma_- - \gamma_+)$
and $(q_+Q)^2=(q_+^2-q_-^2)$.
The function 
$\Theta$ is the Heaviside function, $\Theta(x)$=1 if $x>0$,
and the $I$ functions are modified Bessel functions.
The $\delta$ function terms
carry the probability of the random walker cell having
performed no reversals, and the
Heaviside terms embody causality.

The total particle probability distribution function is
$P(x,t) = P_+(x,t) + P_-(x,t)$.
Under the symmetric condition $a_+=a_-=1/2$, 
and $\gamma_+=\gamma_-$
(or equivalently $q_-=0$) the solution reduces to
that for the generalized Smoluchowski
diffusion equation determined by Hemmer \cite{hemmer},
which at large times is a spreading Gaussian.
The Fourier transform of the distribution functions
yields the moments of the random walker cell
from which the cumulants can be extracted.

\section{Chemokinesis and Chemotaxis}

We now consider the generation and sensing of chemoattactant
or chemorepellant, and its effect on 
the reversal rates ${\gamma_{\pm}}$.
{\it M. xanthus} cells
deposit attractant chemical matter contained within a fibril slime
under the area of cell contact. We assume that 
the release of the chemoattractant
occurs at a constant rate and
uniformly over the cell-substrate footprint.
In general, for a cell of length $\ell$,
the fibril concentration $\phi(x,t)$ obeys the following equation:

\begin{eqnarray}
\label{phi}
\dot \phi(x,t) &=& D \nabla ^2 \phi(x,t) + \\
\nonumber
&& r \Theta[l/2-|x-X(t)|] -r_d \phi(x,t),
\end{eqnarray}

\noindent
where the chemical is generated at a constant rate $r$ per unit area.
The Heaviside function $\Theta$ in Eq.\,(\ref{phi}) 
limits the increase in fibril concentration to the region
strictly under the footprint of the bacterium centered at $X(t)$,
whereas 
degradation and diffusion are included via the terms proportional to
$r_d$ and $D$.
If the surface-deposited chemoattractant does not diffuse
appreciably and $D \nabla ^2 \phi$ is neglected, we find:

\begin{eqnarray}
\label{phitotal}
\phi(x,t) &=& \phi(x,0) e^{-r_d t} +  \\
\nonumber
&&
r \int_0^t dt' e^{r_d (t'-t)}
\Theta[\ell/2-|x-X(t')|]. 
\end{eqnarray}

\noindent
Note that $\phi(x,t)$ depends on the entire history of the
trajectory $X(t')$ up to time $t'=t$.
To couple the fibril concentration with
cell dynamics, 
we must now include a relation between
$\gamma_{\pm}(x,t)$ and
$\phi(x,t)$. 
This relationship will depend on the particular mechanism
of cell direction reversal and various choices will be discussed
in the following subsections.
It is important to note that,
since we are assuming that fibril sensing 
is the main factor in determining direction reversals, 
we are also implicitly assuming that the sensing process, 
for example via a biochemical network within the bacterium,
is much faster than the typical time for a cell
to have moved appreciably.
If we further take the molecular sensing apparatus
to be distributed uniformly at the cell substrate footprint, 
the total chemoattractant $\Phi(X(t), t)$ 
sensed at time $t$
is just the local contribution integrated over the cellular length:

\begin{equation}
\label{phint}
\Phi(X(t), t) = w \int_{X(t)-\ell/2}^{X(t)+\ell/2} \phi(x,t) dx,
\end{equation}

\noindent
where $w$ is the width of the cell as depicted in Fig.\,\ref{1D}.

\subsection{Chemokinesis}
Let us consider the case of chemokinesis
first. Here, the reversal rate depends only
on the integrated fibril concentration. 
We will assume no intrinsic drift and set
$\gamma (x,t) = \gamma_{+}(x,t) = \gamma_{-}(x,t)$.
It is also reasonable to expect 
the switching probability to saturate for large enough fibril concentration,
when the cell cannot respond to further
increases in fibril levels. 
A plausible functional dependence for
$\gamma$ on the total sensed chemoattractant $\Phi$
is of the type found in cooperative chemical binding, 
a Michaelis-Menten form with Hill coefficient $\beta$,
so that for a cell at position $X(t)=x$, the reversal rate is:

\begin{eqnarray}
\label{hill}
\gamma(x,t)&=& \gamma_0 + \kappa(x,t), \\
&=& \gamma_0 + \delta p \frac{\Phi^\beta(x,t)}
{\Phi_0^\beta + \Phi^\beta(x,t)},
\nonumber
\end{eqnarray}

\noindent
where $\Phi_0$ and $\beta$, the inflection parameter
and transition sharpness (or Hill coefficient),
are peculiar to the system being considered.
The effect of chemokinesis is measured by
$\delta p$; with the unbiased choice $\gamma_0 =1/2$, 
$\delta p$ is limited to $-1/2 \le \delta p \le 1/2$.

We must now solve Eqns.\,(\ref{pro1}) and (\ref{pro2}) 
with the above spatial and temporal
dependence for $\gamma(x,t)$, which in turn depends on the
concentration $\Phi(x,t)$ via Eqns.\,(\ref{phitotal}) and (\ref{phint}).
The latter equations carry an explicit dependence on 
the past history of the walker.
Our approach will be to consider the $\gamma_0$ contribution 
of Eq.\,(\ref{hill}) as giving rise to 
Eqns.\,(\ref{pro1}) and (\ref{pro2}) with constant reversal rates,
and find its associated Green's function.
The $\kappa(x,t)$ term in $\gamma(x,t)$ of Eq.\,(\ref{hill}) then
generates non-homogeneous differential equations,
whose solution may be obtained via the known
Green's function.
We will then be able to construct
an iterative process to obtain 
$P(x,t)$ as a function of all
preceeding probabilities $P(x',t'<t)$.
To build the Green's matrix function ${\bf G}(x,t)$,
we note that Eqns.\,(\ref{pro1}) and (\ref{pro2}) can be written as:

\begin{equation}
\label{matform}
{\bf{\dot P}}(x,t) = 
{\bf{M}} (x) ~ {\bf P} (x, t),
\end{equation}

\noindent
where ${\bf P}(x,t) =
 \left[ P_+(x,t), P_-(x,t)\right]^T$ and

\begin{equation}
\label{matrixm}
{\bf M}(x) = \left( 
\begin{array}{cc} 
-v_0 \partial_x -\gamma_+ & \gamma_-\\
\gamma_+ & v_0 \partial_x -\gamma_-
\end{array}
\right).
\end{equation}

\noindent
For the case of generic $\gamma_\pm$, independent
of $(x,t)$, the solution to this equation,
with the initial conditions expressed 
by Eqns.\,(\ref{initial}) is given by 
Eqns.\,(\ref{pplus}) and (\ref{pminus}).
In the case $\gamma_+=\gamma_-=\gamma_0$
we refer to these solutions as
$P_+^{(0)}(x,t)$ and $P_-^{(0)}(x,t)$.
The Green's function for this problem
stems from the modified matrix equation:

\begin{eqnarray}
\label{problem}
\nonumber
{\bf{\dot G}}(x-x',t \ge 0) &=& 
{\bf{M}} (x) ~ {\bf G} (x-x', t) +  \\
&&  {\bf 1} \delta(x-x') \delta(t), \\
{\bf G} (x-x', t<0) &=& 0.
\end{eqnarray}

\noindent
The fact that: 

\begin{equation}
{\bf P}^{(0)} 
(x,t) =
\displaystyle
\left( \begin{array}{c} 
P_+^{(0)} \\ P_-^{(0)}
\end{array} \right)(x,t) =
{\bf G}(x,t) 
\left( \begin{array}{c} 
a_+\\ a_-
\end{array} \right),
\end{equation}

\noindent
leads to the conclusion that the Green's matrix function is given by:

\begin{equation}
{\bf G}(x,t) = 
\left( 
\begin{array}{cc} 
\displaystyle P_+^{(0)} (x,t)_{ a_+=1} & P_+^{(0)}(x,t)_{a_+=0} \\
\displaystyle P_-^{(0)} (x,t)_{ a_-=0} & P_-^{(0)}(x,t)_{a_-=1}
\end{array}
\right).
\end{equation}

\noindent
This result can also be verified by direct evaluation 
of ${\bf G}(k, \omega)$ in Fourier space.
From Eqns.\,($\ref{pro1}$)
and ($\ref{pro2}$) it follows that
reversal rates of the type
$\gamma_\pm (x,t) = \gamma_0 + \kappa_\pm (x,t)$ yield
a new matrix equation:

\begin{equation}
\label{nonhomo}
{\bf \dot P}(x,t) = {\bf M}(x) {\bf P}(x,t) + 
{\bf K}(x,t) {\bf P}(x,t).
\end{equation}

\noindent
where:

\begin{equation}
{\bf K}(x,t) =
\left( 
\begin{array}{cc} 
- \kappa_+(x,t) & \kappa_- (x,t)\\
\kappa_+(x,t)& -\kappa_-(x,t)
\end{array}
\right).
\end{equation}

\noindent
The resulting equation for $P(x,t)$ can be solved iteratively:

\begin{eqnarray}
\label{solution}
{\bf P}(x,t) &=& {\bf P}^{(0)}(x,t)
+ \\
\nonumber
&& \hspace{-0.8cm} \int_{0}^{t} dt'
\int_{-\infty}^{\infty} dx' ~
{\bf G}(x-x',t-t') ~ {\bf K}(x',t') 
~ {\bf P} (x',t').
\end{eqnarray}

\noindent
In the case of chemokinesis, the non-homogeneous
term $\kappa _+(x,t)=\kappa_-(x,t)$ is
contained in Eq.\,(\ref{hill}).
In order to simplify the
expression for $\Phi(x,t)$, 
we assume unit cell width and take the limit
of cellular length $\ell \rightarrow 0$. We also
neglect diffusion and decay of
the fibril.
Under these assumptions, 
$\dot \phi(x,t) = r \ell \delta(x-X(t))$ and
$\Phi(x,t) = w \ell \phi(x,t)$.
The problem is now a deterministic one,
as exemplified by the delta-function in the expression for
$\dot \phi$. In order
to apply Eq.\,(\ref{solution}),
which gives the statistical probability for one bacterium
to lie at $(x,t)$, we need to know its
${\it exact}$ location at $X(t')$ for all previous times.
In other words, the complete history of the cell is
needed to determine future motion and solutions cannot be found.

Nevertheless, we can approximate
$P_\pm (x,t)$ by an averaged density $\rho_\pm$,
utilizing a mean field theory approach.
Here, we average the probabilities 
$P_\pm$ for a single walker over many distinct independent
realizations, so that the self interaction is to 
be expressed on the {\it average} of all 
replicas of the system, and does not depend on the
history of individual walkers.
We can now rewrite the equation for the fibril concentration
as:


\begin{equation}
\label{concentration}
\dot \phi(x,t) 
= r \ell \rho(x,t),
\end{equation}

\noindent
where we indicate the total bacterial density by $\rho(x,t)$.
The above relationship signifies that, on average, 
fibril growth is proportional
to the total density of bacteria at $(x,t)$.
Integrating Eq.\,(\ref{concentration}) and taking
the initial condition 
$\phi(x,0)=0$, we obtain:

\begin{equation}
\label{phint2}
\phi(x,t) = r \ell \int _0 ^t dt' \rho (x,t'),
\end{equation}

%


\noindent
from which, using Eq.\,(\ref{solution}) and the definition
$\Delta \rho(x,t) = \rho_+(x,t)-\rho_-(x,t)$
we can write:

\begin{eqnarray}
\label{sdeltarho}
{\Delta \rho}(x,t) &=& {\Delta \rho}^{(0)}(x,t) +  \\
\nonumber
&& 
\hspace{-1.3cm}
\delta p \int _0 ^t dt' \int_{-\infty}^{\infty}  dx'g (x-x',t-t') 
\frac{\Delta \rho (x',t')\phi^\beta(x',t')} 
{\phi_0^\beta+\phi^\beta(x',t')}. 
\end{eqnarray}

\noindent
Here $g(x,t)$ is a combination of the
matrix elements of the Green's function:
$g= -g_{11}+g_{12}+g_{21}-g_{22}$ and is even in $x$.
An equation  similar to Eq.\,(\ref{sdeltarho}) 
can be obtained for the total
density $\rho(x,t)$, which depends on $\Delta \rho(x,t)$:

\begin{eqnarray}
\label{srho}
{\rho}(x,t) &=& {\rho}^{(0)}(x,t) + \\
\nonumber
&& 
\hspace{-1.3cm}
\delta p
\int _0 ^t dt' \int_{-\infty}^{\infty}  dx'h (x-x',t-t') 
\frac{\Delta \rho (x',t')\phi^\beta(x',t')} 
{\phi_0^\beta+\phi^\beta(x',t')}.
\end{eqnarray}

\noindent
Here, $h=-g_{11}+g_{12}-g_{21}+g_{22}$
and is odd in $x$, as required by the normalization of both
$\rho(x,t)$ and $\rho^0(x,t)$.
It is useful to note that
the above recursive equations may be used to evaluate
the number density $n(x,t)$ for a system of $N$ interacting 
bacteria as well. 
In this case, the distribution function $n_\pm(x,t) 
= N \rho_\pm(x,t) $ and the evolution equations 
are the same as 
($\ref{sdeltarho}$) and ($\ref{srho}$) provided
we redefine $\phi_0 = N n_0$.
Equations $(\ref{phint2})$,
($\ref{sdeltarho}$) and ($\ref{srho}$)
can now be solved numerically.
Once $\Delta \rho (x,t) $ and $\rho (x,t)$ are determined,
$\phi (x,t)$ can be evaluated and used to calculate
the quantities of interest at subsequent times.

\subsection{Chemotaxis}

The incorporation of chemotaxis into the determination of the 
reversal rate 
follows closely that of chemokinesis. 
Let us consider chemotaxis as the 
sole reversal mechanism for a cell of length $\ell$ 
traveling at $v_0$ speed in the positive direction, and situated at
position $X(t)=x$.
The chemical gradient over the cellular length will drive the 
reversal rate. To specify 
the relation between gradient and reversal probabilities, 
we first introduce the probability
terms $\mu_\pm (x,t)$:

\begin{equation}
\label{hopone}
\mu_\pm(x,t)= \exp 
\left[ \mp \sigma \frac{\phi(x+\ell/2, t)-\phi(x-\ell/2, t)}{\ell} \right],
\end{equation}

\noindent
where $\sigma$ is a constant 
that measures the strength of 
chemotaxis.
For positive $\sigma$, the above expression 
translates into a low reversal probability whenever
$\phi(x+\ell/2,t) > \phi(x-\ell/2,t)$ so that the cell is likely
to keep moving speed at $+v_0$. 
This means we are modeling an attractant fibril trail.
On the other hand, negative values of 
$\sigma$ will represent
a chemorepellant system.
The same argument can be presented for a bacterium traveling
at $-v_0$ speed. In this case, the direction of motion tends to
persist for negative gradients, hence the expression for
$\mu_-(x,t)$, where positive $\sigma$ values signify chemoattractant 
systems.
In the limit of cellular length $ \ell \rightarrow 0$,
the above expressions are written as:

\begin{eqnarray}
\label{hone}
\mu_\pm (x,t) = \exp\,[\mp \sigma \phi_x (x,t)]
\end{eqnarray}

\noindent
where the $\phi_x = \partial \phi / \partial x$. We use this exponential
function to capture the sensitive chemical signalling, 
such as that found in {\it E. coli} \cite{segall}.  
Another alternative for the function $\mu_{\pm}$ is to include a saturation 
in the form of $\phi_{x}/\phi$ in the exponent of 
(\ref{hone}).

We assume the reversal 
probabilities $\gamma_\pm (x,t)$ at position $(x,t)$
to depend on $\mu_\pm (x,t)$ through the following:

\begin{equation}
\label{chemorev}
\gamma_\pm (x,t)= \frac 1 2 \left[ 1 + \frac{\mu_\pm (x,t)-1}
{\mu_\pm(x,t) +1} \right],
\end{equation}

\noindent
so that for $\sigma=0$, in the absence of chemotaxis, the
reversal probabilities are $\gamma_\pm =
\gamma_0 = \frac 1 2$.

We are now able to use the same formulation derived for
the chemokinesis mechanism and 
apply it to the chemotactically driven case.
Again, we must resort to the mean field calculation of the
densities $\rho(x,t)$ and $\Delta \rho (x,t)$ by means of the Green's 
matrix function.
The matrix ${\bf K}$ appearing in the 
non-homogeneous term of Eq.\,(\ref{nonhomo}) now has components 
$\kappa_\pm (x,t)$ given by:

\begin{eqnarray}
\label{chemokappa}
\kappa_\pm(x,t) = \frac{\mu_\pm(x,t)-1}{\mu_\pm(x,t)+1},
\end{eqnarray}

\noindent
and Eq.\,(\ref{solution}) still holds.
Note that since $\mu_- (x,t)= 
\mu_+^{-1}(x,t)$, then 
$\kappa_+(x,t)+ \kappa_-(x,t) =0$.
Repeating the same type of calculations as for the chemokinesis case,
for $\Delta \rho$ under chemotaxis we find:

\begin{eqnarray}
\label{sdeltarhochem}
{\Delta \rho}(x,t) &=& {\Delta \rho}^{(0)}(x,t) + \frac 1 2 \\
\nonumber
&& 
\hspace{-1.6cm}
\int _0 ^t dt'
\int_{-\infty}^{\infty} dx'g (x-x',t-t') \rho(x',t') \kappa_+(x',t').
\end{eqnarray}

\noindent
Here the previous history of the cell 
is contained in $\kappa_+(x',t')$
through the derivatives of the fibril 
concentration $\phi(x',t')$. The $g$ 
function is the same as defined in the case of chemokinesis.
Similarly, the density $\rho(x,t)$ is:

\begin{eqnarray}
\label{srhochem}
{\rho}(x,t) &=& {\rho}^{(0)}(x,t) + \frac 1 2 \\
\nonumber
&& 
\hspace{-1.4cm}
\int _0 ^t dt'
\int_{-\infty}^{\infty} dx'h (x-x',t-t') \rho(x',t') \kappa_+(x',t')
\end{eqnarray}

\noindent
and $\rho(x,t)$ is properly normalized by the fact that 
$h(x,t)$ (defined in the preivous subsection)
is odd with respect to $x$.

\subsection{Combined Chemokinesis and Chemotaxis}

If the two mechanisms of
chemokinesis and chemotaxis are active,
both $\delta p$ and $\sigma$ are non-zero, and
$\gamma_\pm (x,t)$ must include both contributions.
For $\gamma_\pm(x,t)$, we write:

\begin{equation}
\label{gp}
\gamma_\pm (x,t) = \frac 1 2 \left\{1+ \frac{p_k (x,t) [\mu_\pm (x,t)+ 1] -1}
{p_k (x,t) [\mu_\pm(x,t) - 1] +1} \right\}.
\end{equation}

\noindent
Here, $p_k(x,t)$ is the contribution from chemokinesis in the form
expressed by Eq.\,(\ref{hill}) with $\gamma_0 = 1/2$, i.e. 
$p_k(x,t) = \gamma(x,t)$ from the chemokinesis case. 
The contribution from chemotaxis is  $\mu_\pm(x,t)$ 
as defined by Eq.\,(\ref{hone}).

The choice $\sigma =0$ sets $\gamma_\pm(x,t) = p_k(x,t)$, that is,
the rate is dictated only by the chemokinesis mechanism.
Conversely, $\delta p = 0$ sets 
$p_k =1/2$ and $\gamma_\pm(x,t)$ reduces to the purely
chemotactic expressions of Eq.\,(\ref{chemorev}).
For both $\sigma = \delta p = 0$ the reversal rate is
$1/2$.
The corresponding $\kappa_+(x,t)$ and $\kappa_-(x,t)$ are contained 
in Eq.\,(\ref{gp}) and yield the following recursive 
relationship for $\Delta \rho (x,t)$:
 
\begin{eqnarray}
\label{sdall}
\nonumber
{\Delta \rho}(x,t) &=& {\Delta \rho}^{(0)}(x,t) +
\int _0 ^t dt' \int_{-\infty}^{\infty}  dx'g (x-x',t-t') \\ 
&& 
\hspace{-0.3cm}
\frac 1 2 
\left[ {\cal U}(x',t') \rho(x',t') + {\cal V} (x',t') \Delta \rho(x',t')
\right].
\end{eqnarray}

\noindent
The corresponding equation for $\rho(x,t)$ is:

\begin{eqnarray}
\label{srall}
\nonumber
{\rho}(x,t) &=& {\rho}^{(0)}(x,t) +
\int _0 ^t dt' 
\int_{-\infty}^{\infty}  dx'h (x-x',t-t') \\ 
&& 
\hspace{-0.3cm}
\frac 1 2 
\left[ {\cal U}(x',t') \rho(x',t') + {\cal V} (x',t') \Delta \rho(x',t')
\right].
\end{eqnarray}

\begin{figure}[t]
\includegraphics[height=3.2 in, angle = -90]{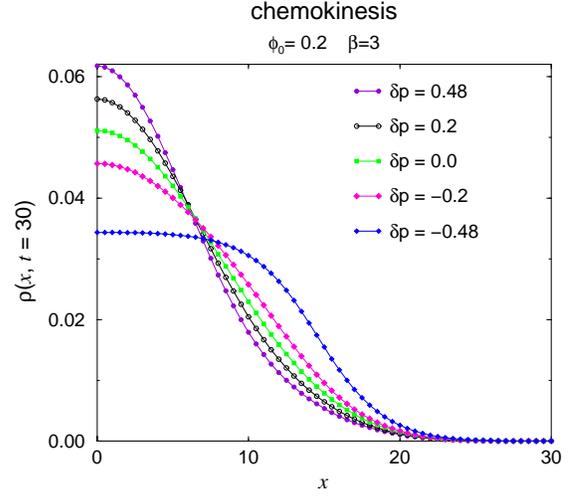}
\vspace{0.2cm}
\caption{
Distribution functions in the case of chemokinesis for $\phi_0=0.02$
and $\beta=3$. Note that $\delta p >0$ increases the reversal rate,
localizing the cell at the initial position and 
$\delta p < 0$ values have the opposite effect.
The time scale is $n=600$ time steps in units of $\Delta t =0.05$.}
\label{kin2}
\end{figure}

\begin{figure}[t]
\includegraphics[height=3.2 in, angle = -90]{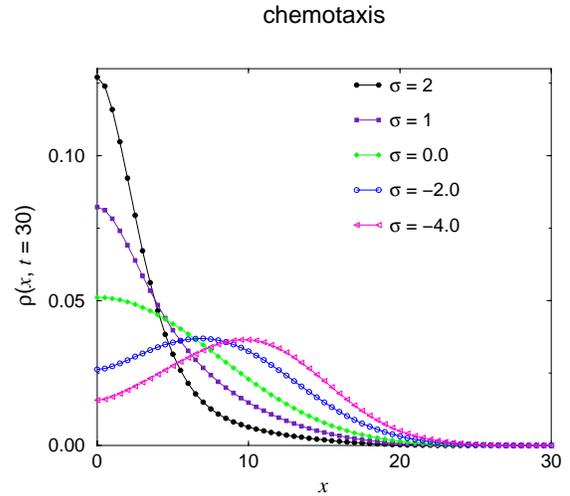}
\vspace{0.2cm}
\caption{
Distribution functions in the case of chemotaxis 
for different $\sigma$ values.
Note that $\sigma >0$ values represent chemoattraction and the cells
are localized at the initial position,
$\sigma < 0$ values instead represent chemorepulsion and two reprelling 
opposite peaks develop.
The time scale is $n=600$ time steps in units of $\Delta t =0.05$.}
\label{chem1}
\end{figure}

\begin{figure}
\includegraphics[height=3.2 in, angle = -90]{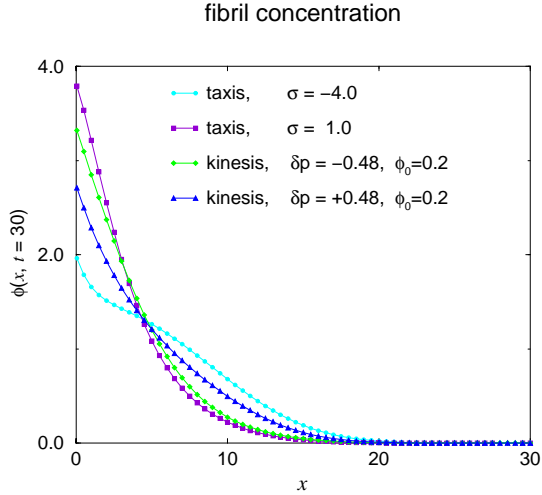}
\vspace{0.2cm}
\caption{
Fibril concentration $\phi(x,t)$ in the case of either 
chemotaxis or chemokinesis
for different $\sigma$ and $\delta p$ values.
The time scale is $n=600$ time steps in units of $\Delta t =0.05$.
In the chemokinesis curves $\phi_0=0.2$, close to the saturation limit of
the Michaelis-Menten form of Eq.\,\ref{hill}.}
\label{chemofibril}
\end{figure}

\noindent
Here, the ${\cal U} (x,t)$ and ${\cal V} (x,t)$ functions
depend on chemotaxis and chemokinesis via:

\begin{eqnarray}
{\cal U}&=& \frac{p_k(1-p_k)(1-\mu_+^2)}
{[p_k(\mu_+-1)+1][\mu_+(p_k-1)-p_k]}, \\
\nonumber
\\
{\cal V}&=& \frac{\mu_+(1-2 p_k)}
{[p_k(\mu_+-1)+1][\mu_+(p_k-1)-p_k]}.
\end{eqnarray}

\noindent
We have used $\mu_+ \mu_-=1$ and suppressed the $(x,t)$ dependence.
Note that in the absence of chemotaxis, for $\sigma=0$
and $\mu_+=1$, ${\cal U} = 0$ and ${\cal V} = 2p_k -1$,
leading to the pure chemokinesis case.
In the absence of chemokinesis, for $p_k=1/2$,
${\cal U}= \kappa_+$, as defined in the previous subsection,
and ${\cal V} = 0$.
We shall discuss the numerical solution to the
equations for $\rho(x,t)$ and $\Delta\rho(x,t)$
in the various cases of chemokinesis, chemotaxis or both,
in the next Section.

\section{Distribution functions}

In this Section, by means of numerical integration,
we solve equations (\ref{srho}), 
(\ref{sdeltarho}),
(\ref{srhochem}),
(\ref{sdeltarhochem}),
(\ref{sdall}), and
(\ref{srall}) for
$\rho(x,t)$ and $\Delta \rho(x,t)$.
Under chemokinesis, Eqns.\,(\ref{srho}) and (\ref{sdeltarho})
are Volterra equations of the second 
kind in the $t$ variable, for which
standard methods can be applied \cite{numrec}.
In particular, 
we discretize both the temporal
and spatial axis according to a uniform mesh,
of spacing $\Delta t$ and $\Delta x$,
and solve the two coupled
equations iteratively.

Let us assume that $\Delta \rho (x', t' < t)$ and $\rho(x', t'< t)$
are known, where $t'$ represent time steps up to $t - \Delta t$.
At the subsequent time step $t$,\, 
$\Delta \rho (x,t)$ is evaluated by calculating 
$\Delta \rho^{(0)}(x,t)$ and by adding to it the sum over all
previous time steps $t'< t$ of the integrand. 
Of these terms, the last one, at $t=t'$,
contains the kernel function $g(x-x',0) = -2\delta (x-x')$,  
which, when integrated over space,
extracts a term proportional to the quantity of interest,
$\Delta \rho (x,t)$.
The other contributions of the sum, for $t \neq t'$, are
spatial integrals, that are evaluated as
sums after the discretization
over the $x'$ variable.
These sums contain the 
known values of $\Delta \rho (x',t'< t)$,
as well as $\kappa(x',t')$ which depends on
the fibril concentration $\phi(x',t')$. 
The latter is just the integrated $\rho(x',t')$
as follows from Eq.\,($\ref{phint2}$).
In order to evaluate $\phi(x',t')$, a time discretized
summation over the $\rho(x',t'<t )$ values is carried out.
The assumption here is that the bacterium
senses the environment before laying down new fibril, so that the
concentration $\phi(x',t')$ is 
evaluated not by the trapeziodal rule,
in which the extrema of the integration region are weighed 
each as 1/2, but by weighing $t=0$ with 1 and ignoring $t=t'$.
In the case of chemokinesis,
once $\Delta \rho(x,t)$ is determined, $\rho(x,t)$,
which depends on $\Delta \rho(x',t')$, can be calculated in the same
way.

In Fig.\,\ref{kin2} 
we plot the results for $\rho(x,t)$
for a system under chemokinesis.
The meshes are set at $\Delta x = \Delta t = 0.05$
and the maximum time step, in these units is $t = 600$.
Cell velocity is fixed at $v_0=1$ and the parameter 
$\phi_0= 0.2$.
For chemoattractant fibril, or $\delta p >0$, 
the distribution is
narrower than the bare $\delta p =0$ case, also plotted in the graph.
Negative values of $\delta p$ represent chemorepulsion 
and broaden the distribution, eventually
leading to a plateau-like region.
The existence of the plateau is due to the fact that for large negative
values of $\delta p \simeq -1/2$ the reversal rate at saturation
is almost zero, leading to ballistic motion of the cell in the region where
the fibril has been deposited. Once the cell reaches the edge of the
plateau region of fibril concentration, 
the reversal rate increases dramatically and the cell tends
to reverse its motion and travel backwards, until the opposite side of the
plateau region is reached. This behavior accounts for the uniformity
of the distribution function in the central region.
Time evolution also tends to broaden - and consequently lower - 
the plateau region, since the
bacteria tend to build up fibril at the edges, thus increasing the 
range of motion.

For the case of chemotaxis,
$\rho(x,t)$, is easier to evaluate,
since it can be calculated without recourse to
$\Delta \rho (x,t)$, as indicated by Eq.\,($\ref{srhochem}$).
The procedure is the same as that outlined for the case of 
chemokinesis.
We take the gradient appearing in the expressions
for $\mu_\pm$ to be evaluated over a distance $\ell = 2$, so that
for $x=x_i$ and $t=t_i$ of the mesh we have:

\begin{equation}
\label{munum}
\mu_\pm (x_i, t_j)= \exp \left[ \mp \frac \sigma 2 [\phi(x_{i+1}, t_j) 
- \phi(x_{i-1}, t_j)] \right].
\end{equation}

\noindent
Results are plotted in Fig.\ref{chem1}.
As discussed in the previous section, 
positive (negative) values of $\sigma$
indicate chemoattraction (chemorepulsion).
It is interesting to note that for large negative values of
$\sigma$ two peaks develop (only one is seen from Fig.\,\ref{chem1},
the other is its symmetrical image in the $x$ axis),
with the bacteria traveling towards unexplored
regions of space, void of fibril.

\begin{figure}[tb]
\includegraphics[height=3.2 in, angle = -90]{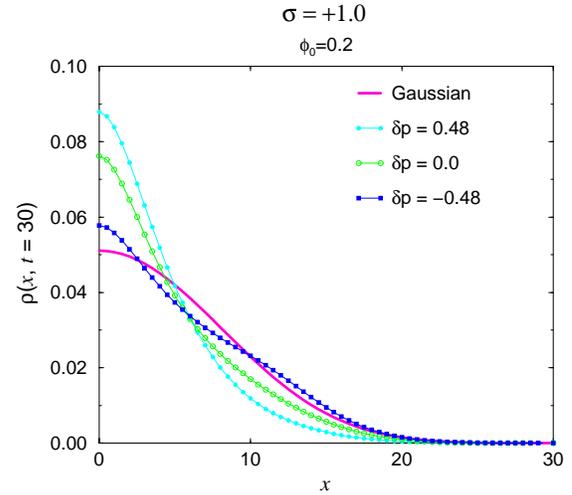}
\vspace{0.5cm}
\caption{
Distribution functions under chemotaxis,  
set at $\sigma=+1.0$, and chemokinesis set at various $\delta p$
values.
The time scale is $n=600$ time steps in units of $\Delta t =0.05$.}
\label{comp1}
\end{figure}

\begin{figure}[tb]
\includegraphics[height=3.2 in, angle = -90]{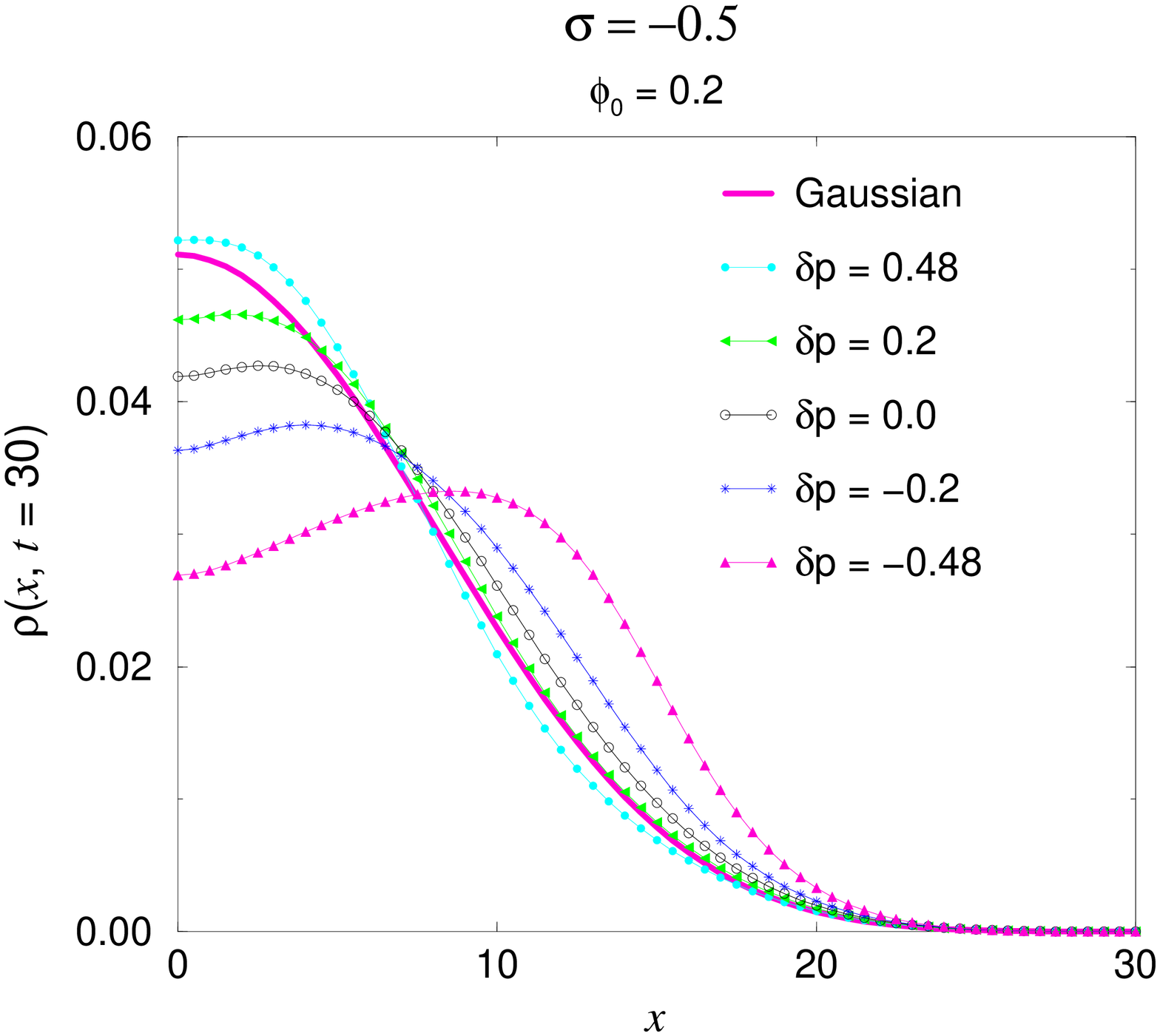}
\vspace{0.5cm}
\caption{
Distribution functions under chemotaxis,  
set at $\sigma=-0.5$, and chemokinesis set at various $\delta p$ values.
The time scale is $n=600$ time steps in units of $\Delta t =0.05$.}
\label{comp3}
\end{figure}

\begin{figure}[tb]
\includegraphics[height=3.2 in, angle = -90]{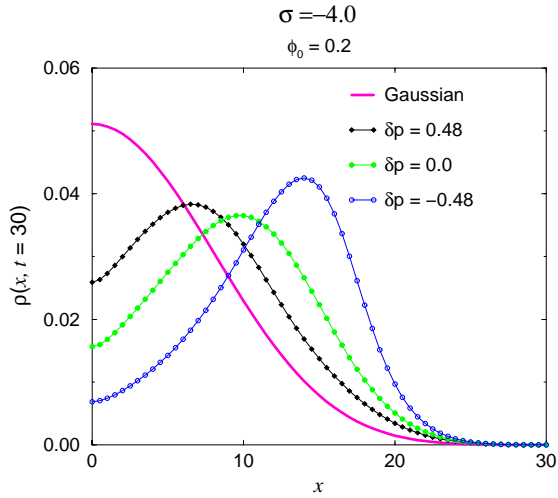}
\vspace{0.5cm}
\caption{
Distribution functions under chemotaxis,  
set at $\sigma=-4.0$, and chemokinesis set at various $\delta p$ values.
The time scale is $n=600$ time steps in units of $\Delta t =0.05$.}
\label{comp5}
\end{figure}

When both mechanisms are present, chemotaxis
and chemokinesis may enforce or compete against each other
in their
localizing or spreading effects.
Chemoattractant values of $\sigma>0$ for instance,
contrast the ballistic tendency of $\delta p <0$, and viceversa,
chemorepulsive values of $\sigma<0$ oppose
the localizing effect of $\delta p > 0$.
In Figures \ref{comp1} through \ref{comp5},
we examine the interplay
of the two mechanisms for several values of $\sigma$
and $\delta p$. 

For both $\sigma,\,\delta p > 0$ the distribution functions 
are localized about the central position, whereas for $\sigma,\,\delta p <0$
two peaks traveling in opposite directions develop. 
The sharpness of these 
peaks increases with increasing
$|\sigma|$, as can be seen by comparing
Figures \ref{comp3} through \ref{comp5}.
and the localization position increases
with increasing $|\delta p|$ as each figure also shows.
Competing effects arise when $\delta p$ and $\sigma$ are 
of opposite signs. For large negative values of $\sigma$,
the effect of positive $\delta p$ values is simply to shift the
mean of the spreading peak towards the origin 
(Figures \ref{comp5}).
For smaller negative values of $\sigma$ increasing
$\delta p$ tends to localize the distribution functions at the origin,
until the traveling peak is washed out and a plateau region,
typical of large positive $\delta p$ values
forms (Fig.\,\ref{comp3}).
Finally, in the regime of $\sigma>0$ and $\delta p<0$,
the localizing tendency due to chemoattraction is offset
by the chemokinesis-plateau tendency. Unusual distribution
function patterns may arise as in the case of $\sigma=0.5$ or $\sigma=1$
and $\delta p=-0.4$ or $\delta p=-0.48$
in Fig.\,\ref{comp1}.

\section{Monte-Carlo simulations}

To analyze the statistics of the one dimensional
dynamics of cell motion under chemotaxis and chemokinesis,
we also implemented a stochastic simulation on a 1D lattice. 
We discretize space and time and assume the 
random walker cell to have finite length $\ell$.
At time $t=0$, the single cell is positioned at the
origin of a one dimensional track and there is no
fibril substance present. Fibril is constantly
secreted by the cell and the dynamics of the particle position 
obeys:

\begin{equation}
X_{t+1}= X_t + \eta_t (X_t-X_{t-1}),
\end{equation}

\noindent
where $X_{t+1}$ is the position of the center
of the cell at time $t+1$ and 
$X_t-X_{t-1}$ indicates the direction it was traveling prior to
time $t$. 
The stochastic
random variable $\eta_t$ is defined as:

\begin{eqnarray}
\eta_{t}= \left\{ \begin{array}{ll} -1 & \quad \mbox{with prob.} 
\,\,\gamma_t \\[13pt]
+1 & \quad \mbox{with prob.} \,\, 1-\gamma_t \end{array}  \right. \\
\nonumber
\end{eqnarray}

\noindent
where $\gamma_t$  is the probability for the cell 
to reverse its direction, before the time step at $t+1$.
The initial direction of motion
is chosen so that $X_{-1}$ is $\pm 1$ with equal probability.
We simultaneously model chemotaxis and chemokinesis through the 
dependence of $\gamma_t$ on 
the self secreted fibril concentration $\phi(x,t)$,
as in the continuum case.
Discretizing equations (\ref{phi}) and (\ref{phint}) we obtain:

\begin{eqnarray}
\phi_t(x) &=& \sum_{s=0}^t \Theta(\ell/2 - |x - X_s|),\ \mbox{and} \\
\Phi_t(x)&=& \sum_{x=-\ell/2}^{\ell/2} \phi_t(X_t + x),
\end{eqnarray}

\noindent
where $\phi_t(x)$ is the amount of chemoattractant that has built up at 
location $x$ and $\Phi_t$ is the total amount of chemoattractant sensed 
by the cell. 
Both $\phi_t(x)$ and $\Phi_t$ naturally depend on the past 
trajectory of the cell, which we keep track of as the simulation proceeds.
We let $p_{t,k}$ represent the probability of direction reversal based 
on chemokinesis alone derived from equation 
(\ref{hill}).  
This probability is then stretched towards 0 or 1 
depending on the local normalized chemical gradient in chemotaxis, 
through the quantity $\mu_{t,c}$. 
Analogous to Eqs. (\ref{hill}) 
and (\ref{hopone})
in the continuum theory, 
the kinesis and chemotaxis effects are represented by:

\begin{eqnarray}
\label{P}
p_{t,k} &=& \frac 1 2 + \delta p \frac{\Phi_t^{\beta}}
{\Phi_{0}^{\beta}+\Phi_t^{\beta}}, \\
\mu_{t,c} &=& \exp \left[ -\sigma (X_t - X_{t-1}) 
\nabla \phi_t \right],
\end{eqnarray}

\noindent
where $\nabla \phi_t = \phi_t(X_t + \ell/2) - \phi_t(X_t - \ell/2)$ 
and $\delta p$ and $\sigma$ are weight parameters
subject to the constraints:

\begin{equation}
-\frac 1 2  \le \delta p \le \frac 1 2  \mbox{ and }
-\infty < \sigma < \infty.
\end{equation}

\noindent
The parameter $\sigma$ scales the effect size of chemotaxis, 
with no chemotaxis corresponding to $\sigma = 0$; while $\delta p$ 
scales the strength of chemokinesis, with no chemokinesis corresponding 
to $\delta p = 0$.
At each step of the Monte-Carlo iteration, the total 
reversal probability is calculated as:

\begin{equation}
\gamma_t = \frac 1 2 \left[ 1 +  
\frac{p_{t,k}(\mu_{t,c}+1) -1}
{p_{t,k} (\mu_{t,c}-1) + 1} \right].
\end{equation}

\begin{figure}
\includegraphics[height=3.2 in, angle=-90]{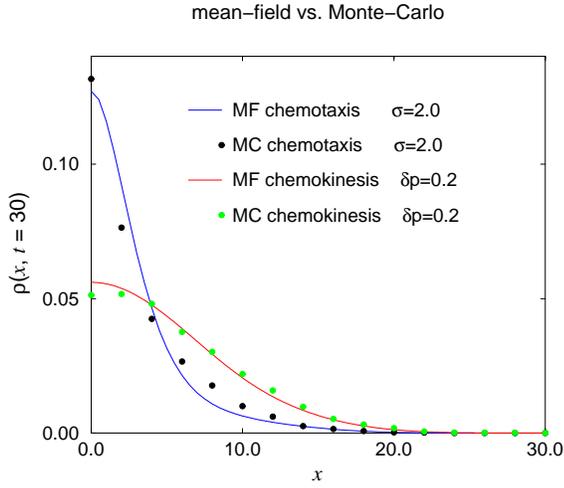}
\vspace{0.5cm}
\caption{
Comparison of mean-field theory results with Monte-Carlo simulations
for short times. The mean-field curves are evaluated after $n=600$
time steps of $\Delta t =0.05$ and the Monte-Carlo curves are
determined after $t=30$ time steps. In the chemokinesis Monte-Carlo
curve $\Phi_0 = 4$, whereas in the chemokinesis mean-field
curve $\phi_0 = 0.2$. The factor of $0.05$ is needed 
to take into account the discretization of the numerical Volterra-type
evaluation.}
\label{compare}
\end{figure}

\noindent
In Fig.\,\ref{compare} we compare the results from the 
Monte Carlo simulations and the mean-field approximation of
the previous section, for small times, for a typical case
of chemokinesis and chemotaxis.

\begin{figure}
\includegraphics[height=3.2 in, angle=-90]{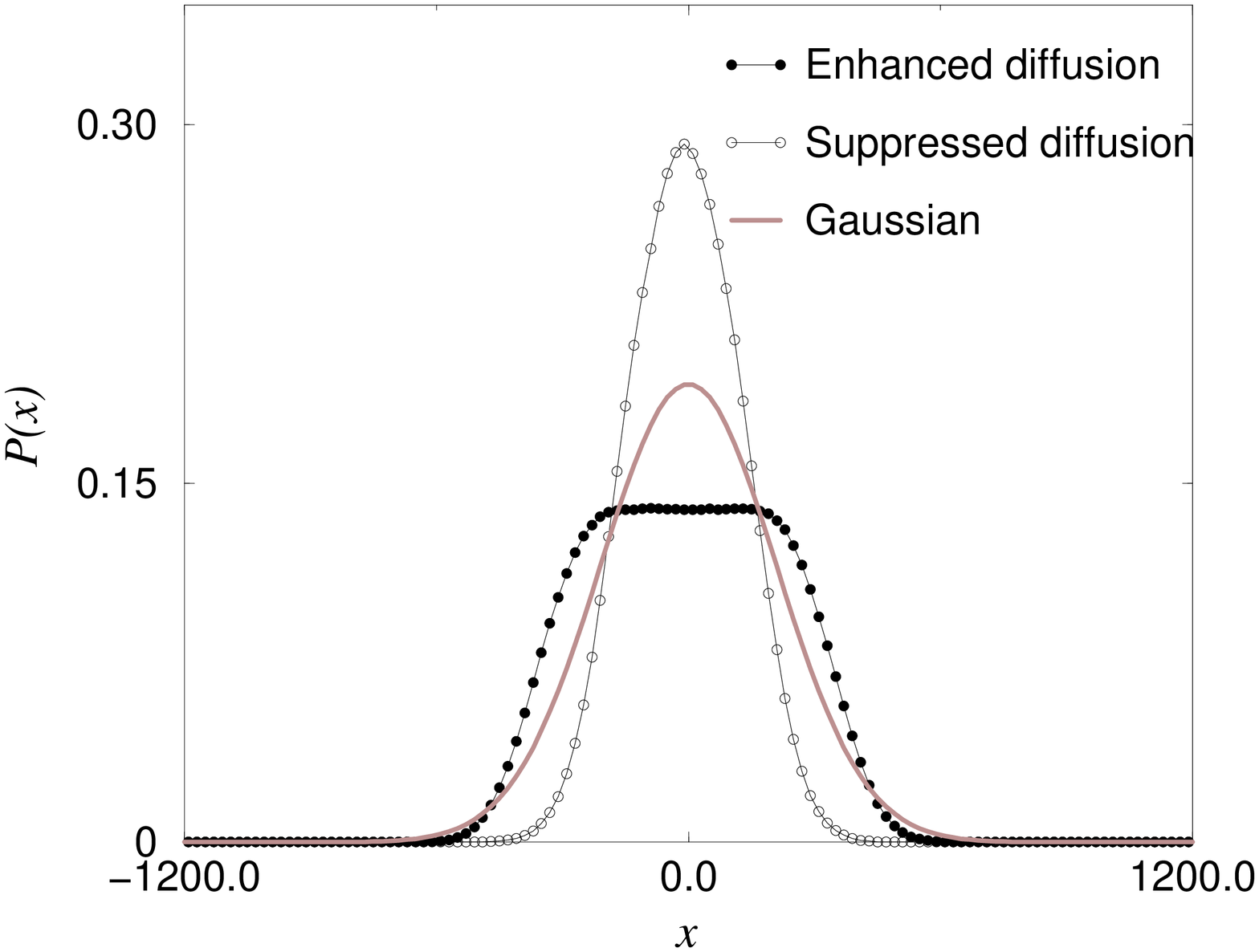}
\includegraphics[height=3.2 in, angle=-90]{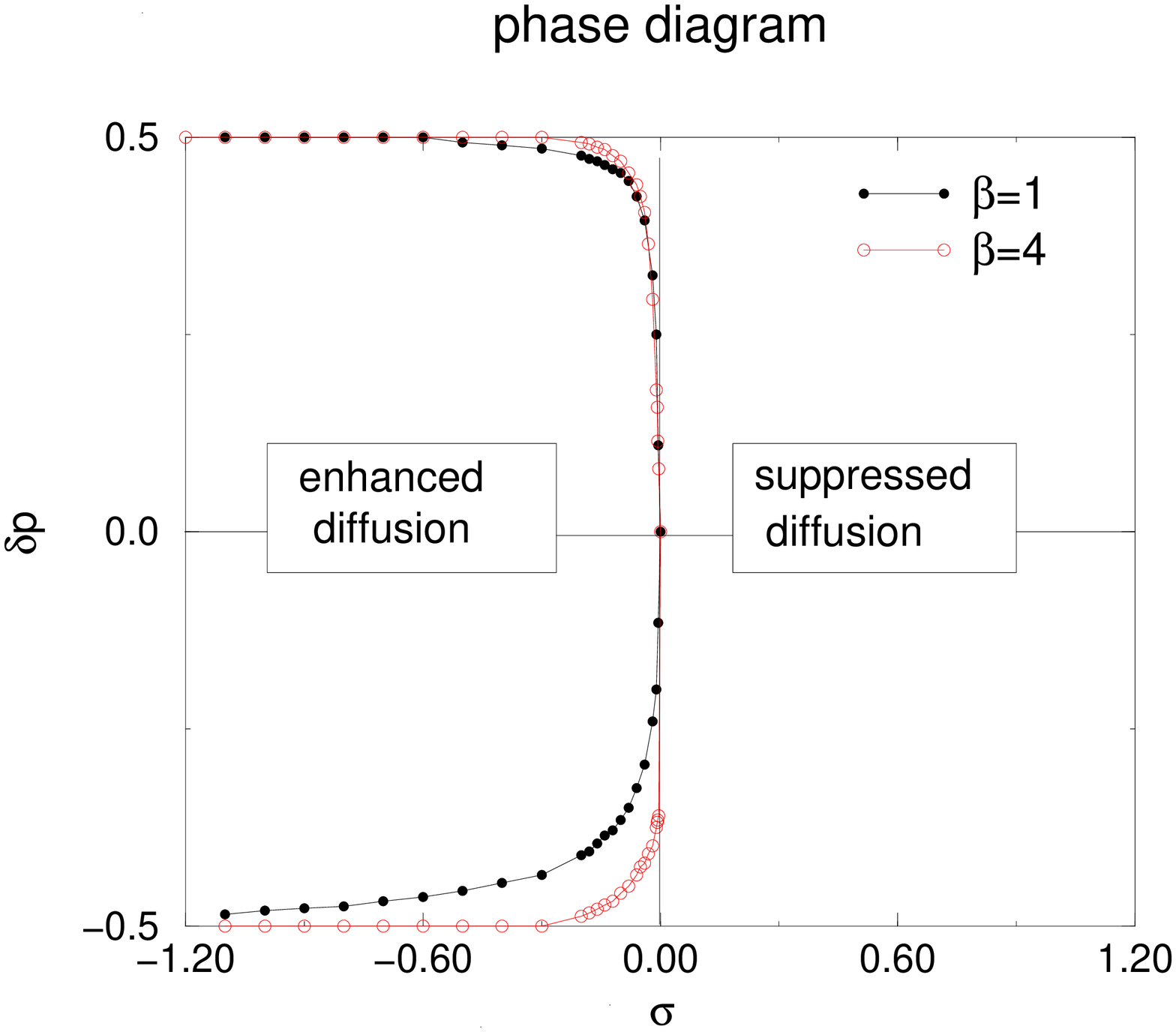}
\vspace{0.5cm}
\caption{(top) Gaussian and suppressed and enhanced PDFs. (bottom)
$\sigma-\delta p$ phase diagram at $t=10^{5}$
for $\beta = 1, 4$. 
The crossover between suppressed and enhanced diffusion 
was found by Monte-Carlo simulations and determining the point at which 
the integral of the reference Gaussian 
between the curvature inflexion points is equal
to that of the distribution functions.}
\label{phase}
\end{figure}

We can now analyze results of the Monte-Carlo simulations for 
longer times than in the analytic case: $t_{max}=10^5$.
At these time scales, the general 
trends found in the analytic case
persist for most combinations of $(\sigma, \delta p)$.
The simulation results are summarized in a phase-diagram.  Positive values of
$\delta p$ and $\sigma$ both cause attractiveness in the system and we
expect suppressed spreading of the 
distribution function. Negative values of $\delta p$ and
$\sigma$ on the other hand cause repulsion and enhanced spreading.
To quantify these effects, we compared the distribution function 
$P(x,t_{max})$ with a normalized Gaussian, corresponding to the
parameters  $\sigma = \delta p =0$.
More specifically, 
we compare the integral of the reference Gaussian at the
curvature inflexion points to the integral of the
probability distribution function between the same limits.
If the integral of the PDF is
less than that of the corresponding Gaussian, 
the dynamics is classified as being suppressed. 
This definition
is shown in Fig.\,\ref{phase}(top).
We explore the entire phase space $\delta p$ and
$\sigma$ and find the crossover from one regime to the other. 
The enhanced or suppressed diffusive character of the PDF
curves may evolve from that of the early time regime,
but it is found that after $t=10^3$ simulation time steps 
the classification is stable.
The phase diagram is shown in
Fig.\,\ref{phase}(bottom).
For $\sigma, \delta p > 0$, both kinesis and chemotaxis effects
suppress the diffusivity of the trajectories. Only for $\sigma < 0$
can the spreading dynamics be enhanced. Due to our choice of the
functional forms of the effects (Eqs. \ref{hill} and \ref{hopone}),
the dynamics are much more sensitive to $\sigma$ that $\delta p$,
nonetheless, the effects of kinesis induces an asymmetry in $\delta p$
in the phase diagram.  The curve approaches 
the limits $\delta p =\pm 0.5$ asymptotically.

\section{Conclusion}

In this paper we presented a mean field model and Monte-Carlo
simulations for bacterial dynamics under the mechanisms of
chemokinesis and chemotaxis acting concurrently.  The bacterial motion
is that of a one-dimensional self-interacting random walker.
The fibril trail that governs the dynamics of the system is itself
a dynamical quantity that depends on the past motion of the cells.
We model the two mechanisms in terms of two characterizing parameters,
$\sigma$ and $\delta p$, which represent, respectively, the degree of
chemotaxis and chemokinesis.  The mean-field results agree with
Monte-Carlo simulations in the limit of short times. For long
times we find a phase-diagram in ($\sigma$ - $\delta p$) space that
separates enhanced or suppressed diffusion regimes.  In contrast to
the short-time numerics displayed in 
Figs.\,\ref{kin2}-\ref{comp5}, which show PDFs 
indicative of enhanced diffusion for $\delta p<0$, the long time
phase diagram exhibits suppressed diffusion provided $\sigma$ is
negative and $\vert \delta p\vert$ is not too large. 
Therefore, in the long-time limit,
kinesis modeled by the logistic saturation, Eq.\,(\ref{hill}), 
results in
suppressed diffusion in the absence of chemorepellent effects 
($\sigma < 0$). 
We have investigated the effects of 
increased cooperativity in the
biochemical signaling through the $\beta$
parameter.
As can be seen from 
Fig.\ref{phase} increased 
$\beta$ values do not 
significantly affect the
phase-diagram curve.

The model exhibits variations in shapes of the
particle PDF. The 
characteristic PDF shapes depend on whether long or short
time dynamics are considered.  

One of the possible applications of this work is for isolated
A-motility (adventurous) Myxobacteria cells whose dynamics 
is driven by self-secreted slime and does
not require the presence of neighboring cells
in direct contact \cite{oster1, oster2}.
Our analysis is restricted to a one-dimensional 
system. For a single cell constrained to move along a 1D agar track,  
the relevant parameters ($\sigma$, $\delta p$ and $\beta$) of our model
can be tuned to identify the correct mechanism of motion.
Moreover, the motion of freely moving A-type 
myxobacteria cells 
is \emph{locally} one-dimensional.
Our model does not account for contact interactions between cells,
rather, cells interact indirectly via trails of self-secreted slime.   
This simplification enables us to study the details
of the effects of 
the history-dependent fibril concentration. 
Direct cell-cell signalling interactions
on a 2D aggregating colony are known to lead to 
the formation of complex structures 
such as propagating rippling waves and spiraling fruiting bodies
\cite{oster1, tieman}. 
Although the investigation of contact interactions between cells
is beyond the scope of this paper,
our approach does not preclude the 
formation of propagating waves and associated
rippling patterns in systems 
where fibril-mediated cell-cell interactions are included.

The authors are grateful for the assistance of G. Lakatos 
in improving the implementation of our simulations and numerical computations.
MD and TC were supported by the National Science Foundation 
through grant DMS-0206733.




\newpage

\begin{references}

\bibitem{preston} T.M. Preston, C.A. King, and J.S. Hyams,
\textit{The cytoskeleton and Cell Motility}, 
(Blackie, Glasgow and London, 1990).

\bibitem{levine} E. Ben-Jacob, I. Cohen and H. Levine
\textit{Adv. Phys.}, {\bf 49}, 395, (2000). 

\bibitem{berg} H.C. Berg and E.M. Purcell, 
{\it Biophys. J.}, {\bf 20}, 193, (1977).

\bibitem{woodward} D.E. Woodward, R. Tyson, M.R. Myerscough, J.D. Murray,
E.O. Budrene, and H.C. Berg,
{\it Biophys. J.} {\bf 68}, 218, (1989).

\bibitem{bergold} H. C. Berg and D. A. Brown, {\it Nature}, {\bf 239}, 500-504,
(1972).

\bibitem{brenner}
M. P. Brenner, L. S. Levitov and E. O. Budrene,
{\it Biophys. J.} {\bf 74}, 1677, (1998).

\bibitem{kearns} D.B. Kearns and L.J. Shimkets,
{\it PNAS} {\bf 95}, 11957, (1998).

\bibitem{rappel} W. J. Rappel, P. J. Thomas, H. Levine and W. F. Loomis
{\it Biophys. J.} {\bf 83},(2002).

\bibitem{shi} W. Shi and D.B. Kearns,
{\it Nature} {\bf 366}, 414, (1993).

\bibitem{tieman} S. Tieman, A. Koch and D.A. Kaiser,
{\it J. Bacteriol.} {\bf 178}, 3480, (1996).

\bibitem{kearns2} D.B. Kearns, A. Venot, P.J. Bonner, B. Stevens, G. Boons
and L.J. Shimkets,
{\it PNAS} {\bf 98}, 13990, (2001).

\bibitem{Shi96} W. Shi, F. K. Ngok, and D. Zusman, 
{\it Proc. Natl. Acad.
Sci.}, {\bf 93}, 4142-4146, (1996).

\bibitem{soll}
D. R. Soll {\it J. Chem. Ecol.}, {\bf 16}, 1, (1990).

\bibitem{spormann}
A. M. Spormann, {\it Microbiol. Mol. Biol. Rev.} {\bf 63},
621, (1999).

\bibitem{Mer00} A. J. Merz, M. So, and M. P. Sheetz, 
{\it Nature}, {\bf 407}, 98-102, (2000).

\bibitem{zusman} 
M.J. Ward and D.R. Zusman
{\it Mol. Microbiol.} {\bf 24}, 885, (1997).

\bibitem{zusman2} 
M.J. Ward, K.C. Mok and D.R. Zusman
{\it J. Bacteriol.} {\bf 180}, 440, (1998).

\bibitem{Alt}
{\it J. Math. Biol.} {\bf 9}, 147-177, (1980).

\bibitem{Kel74} E. F. Keller, 
{\it Antibiotics and Chemotherapy}, {\bf 19}, 79-93, (1974).

\bibitem{Dal98} J. C. Dallon and H. G. Othmer, 
{\it J. Theor. Biol.}, {\bf 194}, 461-483, (1998).

\bibitem{Pal00} E. Palsson and H. G. Othmer, 
{\it Proc. Natl. Acad. Sci.}, {\bf 97}, 1044810453, (2000).

\bibitem{Nos76} R. Nossal, 
{\it Math. Biosci.}, {\bf 31}, 121-129, (1976).

\bibitem{Oth97} H. G. Othmer and A. Stevens, 
{\it SIAM J. Appl. Math.}, {\bf 57}, 1044-1081, (1997).

\bibitem{hillen} H. G. Othmer and T. Hillen, 
{\it SIAM J. Appl. Math.}, {\bf 57}, 1044-1081, (1997).


\bibitem{Ste95} A. Stevens, 
{\it J. Biol. Sys.}, 
{\bf 3}, 1059-1068, (1995).

\bibitem{rohde} T. Hillen, C. Rohde and F. Lutscher, 
{\it J. Math  Anal. Appl.}, 
{\bf 260}, 173, (2001).

\bibitem{ants}
D. Chowdhury, V. Guttal, K. Nishinari, A. Schadschneider, 
J. Phys. A, \textbf{L573}, (2002).

\bibitem{maritan}
F. Cecconi, M. Marsili, J. R.
Banavar and A. Maritan, Phys. Rev. Lett., \textbf{89}, 2002.

\bibitem{hemmer} P.C. Hemmer, {\it Physica} {\bf 27}, 79-82,
(1961).

\bibitem{segall} J. E. Segall, S. M. Block, and H. C. berg, 
{\it Proc. Natl. Acad. Sci.}, {\bf 83}, 8987-8991, (1986).

\bibitem{numrec} W.H. Press, S.A. Teukolosky, W.T. Vetterling
and B.P. Flannery, 
\textit{Numerical Recipes in C, Second Edition}, Chapter 18.2,  
(Cambridge University Press, Cambridge,  1999).



\bibitem{oster1}
C. Wolgemuth, E. Hoiczyk, D. Kaiser and G. Oster,
\textit{Curr. Biol.} {\bf 12}, 369-377, (2002).

\bibitem{oster2}
O. Igoshin and G. Oster, 
\textit{Math. Biosci.} (in press)





\end{references}

\end{document}